# Erbium-doped-fiber-based broad visible range frequency comb with a 30 GHz mode spacing for astronomical applications

Keisuke Nakamura*, Ken Kashiwagi*, Sho Okubo**, and Hajime Inaba

National Metrology Institute of Japan (NMIJ), National Institute of Advanced Industrial Science and Technology (AIST), 1-1-1 Umezono, Tsukuba 305-8563, Japan

*These two authors contributed equally to this study.
**e-mail: sho-ookubo@aist.go.jp

## Abstract

Optical frequency combs have the potential to improve the precision of the radial velocity measurement of celestial bodies, leading to breakthroughs in such fields as exoplanet exploration. For these purposes, the comb must have a broad spectral coverage in the visible wavelength region, a wide mode spacing that can be resolved with a high dispersion spectrograph, and sufficient robustness to operate for long periods even in remote locations. We have realized a comb system with a 30 GHz mode spacing, 62 % available wavelength coverage in the visible region, and 40 dB spectral contrast by combining a robust erbium-doped-fiber-based femtosecond laser, mode filtering with newly designed optical cavities, and broadband-visible-range comb generation using a chirped periodically-poled $LiNbO_3$ ridge waveguide. The system durability and reliability are also promising because of the stable spectrum, which is due to the use of almost all polarization-maintaining fiber optics, moderate optical power, and good frequency repeatability obtained with a wavelength-stabilized laser.

## Introduction

Optical frequency combs with a wide mode spacing have a high power per mode, and each mode can be resolved with diffraction gratings or optical filters, which is essential for applications such as mode-resolved direct frequency comb spectroscopy [1, 2] and line-by-line arbitrary optical waveform synthesis [3]. In particular, the wavelength calibration of astronomical spectrographs is expected to lead to breakthroughs in exoplanet exploration and cosmological research by improving the precision of radial velocity (RV) measurement [4, 5]. RV measurement using the Doppler shift of the stellar spectrum is known as the "Doppler method" [6] and was used in the discovery of the first exoplanet [7]. A precision of a few cm/s is needed to find earth-like exoplanets with the Doppler method, and this is difficult to achieve with conventional wavelength standards such as Th-Ar lamps and iodine cells. To cope with this situation, an optical frequency comb or "astro-comb," has been proposed and developed as the wavelength standard for RV measurement [8-28].

Astro-combs require a mode spacing several times wider than a resolution of a high-dispersion spectrograph (> 10 GHz) and broad spectral coverage that depends on the celestial body being observed. A frequency comb with a wide spacing needs a high average power to obtain the pulse energy required for spectral broadening due to nonlinear optical effects. In addition, the development of astro-combs is made more difficult because they must be robust and durable for long-term remote operation at observatories. The scheme frequently employed for astro-combs involves increasing the sub-GHz mode spacing of a mode-locked laser to more than 10 GHz using mode-filtering cavities [9-12]. The advantages of this scheme are that it is relatively easy to achieve self-referencing and obtain 100 fs-level optical pulses. The difficulty is that unnecessary modes attenuated by the optical cavities are revived through the spectral broadening process [13, 14]. On the other hand, there have been reports of astro-combs generated by modulating a CW laser with electro-optic modulators [15-18] and by the Kerr effect in a micro-cavity [19, 20] as ways of directly realizing combs with a mode spacing exceeding 10 GHz but without mode-filtering cavities.

The spectral range of the astro-comb is also important. The wavelength most often used in the RV measurement of celestial bodies is the visible region where there are abundant atomic absorption lines. Ytterbium (Yb) doped-fiber-laser-based [21, 22], and the titanium-sapphire-laser-based [23-25] combs have been reported as schemes for obtaining visible broadband astro-combs, because of their short wavelength and high output power. In particular, Yb-fiber-based astro-combs have produced actual results, and combs have been reported with a mode spacing of 18 GHz or 25 GHz and a wavelength coverage of 455 nm-691 nm [26]. This coverage reaches 48 % of the visible region, and RV measurement precision at the 1 cm/s level has been reported. Note that we define the visible wavelength region as 360 nm-830 nm in this paper.

In this paper, we describe an astro-comb scheme that combines a robust erbium (Er) comb, mode-filtering cavities, spectral broadening with highly nonlinear fiber (HNLF) and multi-order harmonic generation with a chirped periodically-poled lithium-niobate waveguide (cPPLN-WG), and a wavelength-stabilized laser. We have achieved unprecedented spectral coverage while ensuring a sufficient mode-spacing frequency and unnecessary-mode suppression ratio (UMSR). We also discuss the possibility of spectral extension to all visible wavelengths.

## Results

### Overview of comb system

Figure 1a shows an overview of the broadband, visible, and wide-mode-spacing comb system. We employ an Er-doped-fiber-based mode-locked laser as the comb source. Its carrier-envelope offset frequency ($f_{CEO}$) and repetition frequency ($f_{rep}$) were phase-locked to reference frequencies from an atomic clock. Using three Fabry-Perot cavities, the comb mode-spacing was increased to 30 GHz by matching one of every 130 modes of the comb to the cavity transmission mode frequency and suppressing the power of other unnecessary comb modes. Here, an acetylene-stabilized laser [29] was used as a reference for cavity-length stabilization so that the comb and cavity-transmission mode-frequencies are easily reproduced. The comb was amplified with two polarization-maintaining Er-doped fiber amplifiers (EDFAs) inserted between the cavities, and then the comb spectrum was broadened in the infrared region with a polarization-maintaining HNLF [30]. We then input the broadband comb into a cPPLN-WG; the second to fourth-order harmonic generation processes converted the comb in the infrared region into a broadband comb in the visible region. This is an evolution of the previous high-order harmonic generation of frequency combs [31-33]. For details of each part, see Methods.

Figure 1b shows the frequency relationship between the comb modes, the transmission modes of the mode-filtering cavities, and the wavelength-stabilized laser. The $f_{rep}$ and $f_{CEO}$ of the comb source were stabilized at 230 MHz and 30 MHz, respectively. The wavelength-stabilized laser was frequency-stabilized to a $^{13}C_2H_2$ absorption line at a wavelength of 1542 nm ($v_1 + v_3$ P(16)); not referring to the comb. The output of the laser wave was divided into two parts; one of which was used for beat detection with the comb. The other was frequency-shifted with an acousto-optic modulator (AOM) and used for the length stabilization of mode-filtering cavities. Here $f_{CEO}$ and $f_{rep}$ were set so that the beat frequency ($f_{beat}$) between the wavelength-stabilized laser and the nearest comb mode was approximately 40 MHz. The laser frequency output from the AOM was feed-forward controlled [34] to match one of the comb mode frequencies by applying the $f_{beat}$ signal to the AOM. This frequency-controlled laser was used as a reference laser for the three mode-filtering cavities. The cavity lengths were controlled to allow the reference laser to transmit. We set the free spectral ranges (FSRs) of the three cavities at approximately 2.00 GHz ($130f_{rep}/15$), 1.77 GHz ($130f_{rep}/17$), and 2.14 GHz ($130f_{rep}/14$), respectively. The combination of these FSRs resulted in a high UMSR after the comb had passed through the three cavities connected in series. The resultant transmitted comb mode spacing was approximately 30 GHz ($130f_{rep}$), and the calculated UMSR was more than 60 dB at a finesse of 100 (see Fig. 3). During the cavity locking procedure, we scanned the optical cavity length and observed the transmitted reference laser power and the total transmitted comb power simultaneously; we locked the cavity mode with the maximum total transmitted comb power to the reference laser wavelength. Thus, the FSR of the optical cavity was closest to the rational multiple of $f_{rep}$; a high transmittance over a broad spectral region was obtained for the extracted comb modes.

**Spectral range**

We employed chirp-pulse amplification [35] to obtain optical pulses with sufficient peak power for spectral broadening using an HNLF. As shown in Fig. 1a (4), 70 % of the output from the mode-locked laser was first highly chirped using a 15-m-long normal-dispersion fiber (NDF), and then passed through three cavities and two EDFAs in the order shown in the figure. The temporal pulse width stretched with the NDF was gradually compressed with anomalous-dispersion fibers used in the mode-filtering and amplification parts. Then, the temporal width of the pulses was compressed so that it was chirp-free by using an anomalous-dispersion polarization-maintaining single-mode fiber (PM-SMF) after the third cavity. The temporal width of the pulse measured with frequency-resolved optical gating (FROG) was 180 fs. From the average power of 880 mW, the peak power was estimated to be 0.13 kW. The compressed 30 GHz-repetitive optical pulse train was incident into the HNLF to broaden the spectrum in the near-infrared region.

Figure 2a shows the spectra of the 30 GHz pulse trains at the HNLF input (solid black line) and output (solid red line). The spectrum at the output of the HNLF calculated with the split-step Fourier method is also shown in Fig. 2a (dashed blue line). The measured and calculated spectra were in good agreement. See Methods for details of the simulation. Figure 2b shows comb-resolved spectra at 1550 nm observed with a high-resolution optical spectrum analyzer (OSA) at the HNLF input (black line) and output (red line). The contrast was approximately 55 dB at the input and 40 dB at the output. Thus, we observed that the spectral contrast at the HNLF output was lower than that at the input. The spectra were also observed in the 1350 nm-1400 nm, 1525 nm-1625 nm, and 1650 nm-1700 nm ranges, where similar mode-resolved comb spectra were observed. The spectral contrast is discussed in detail in the next subsection.

The broadband comb in the near-infrared region broadened with the HNLF was incident in a 10 cm-long cPPLN-WG to generate the second to fourth-order harmonics; the infrared comb was converted into broadband combs in the visible range. The cPPLN-WG had a poling period that varied linearly from 12.8 μm to 19.4 μm and was designed to satisfy the quasi-phase-matching condition of the second harmonic generation from the wavelength range 1350 nm-1600 nm to 675 nm-800 nm. The average power incident into the cPPLN-WG was 700 mW; the power per mode of the incident comb exceeded 10 μW in the design wavelength range of the cPPLN-WG.

Figure 2c shows the spectrum of the broadband comb output from the cPPLN-WG observed with an OSA (solid green line) through a multi-mode fiber (core diameter: ~50 μm). Figure 2d shows the spectrum at a wavelength of 800 nm with high resolution that we observed in the same way, and we obtained well-resolved comb modes. The CCD image sensor used in the high-dispersion spectrograph in which the comb system will be installed begins to saturate when the number of photons detected per pixel reaches $10^5$ [36]. Considering the pixel area, the imaging area of the comb, quantum efficiency of the sensor and optical coupling, the signal begins to saturate when the photon number of the comb mode reaches $10^8$. We assume that the spectrograph can use a comb mode as a wavelength reference if the mode has 1/10 of the saturation photon number at an exposure time of 1 s. In other words, we defined the available wavelength range in which we can obtained $10^7$ photons/(s mode). Then, the available harmonic component ranges are 664 nm-873 nm, 453 nm-543 nm, and 350 nm-408 nm, as shown in Fig. 2c. The available wavelength coverage of the obtained comb reaches approximately 62 % of the visible wavelength region in the frequency domain when the visible region is defined as 360 nm-830 nm. This is the best coverage for a visible-range comb with a mode spacing in the 30 GHz class.

**Spectral contrast**

For precise RV measurement with high-dispersion spectrograph, the imaged comb-spectrum must have a high contrast. The spectral contrast is primarily determined by the quantity of amplified spontaneous-emission (ASE) and the UMSR; the effect of ASE is not negligible for spectral observation with a spectrograph due to its wide resolution bandwidth. In this study, we assume that the quantity of ASE is sufficiently suppressed in the visible region because the final optical cavity is placed after the final EDFA, and the wavelength conversion of the ultra-short optical pulses by nonlinear optical effects acts as filter in the frequency and time domains of the ASE, respectively. Therefore, we considered that the UMSR was the dominant factor determining the contrast. We measured the UMSR of the comb at the HNLF input, the HNLF output, and the cPPLN-WG output using a CW laser as a probe. For details of the measurement procedures, see Methods.

Figure 3 shows the UMSR of the comb at the HNLF input (1542 nm, open blue circles), the HNLF output (1542 nm, filled red circles), and the cPPLN-WG output (514 nm, green diamonds). The light blue line shows the UMSR of the comb output from three mode-filtering cavities. This is almost equivalent to the comb at the HNLF input and can be calculated from the ratios of the cavity FSRs to the comb $f_{rep}$ and the finesse of the cavities (~100). When the order of a certain transmitted mode is considered to be zero, the comb modes with orders of integer multiples of 130 are transmitted modes, and the others are unnecessary modes. In each wavelength range, the UMSR can be determined by measuring the suppression ratio from mode order 0 to 65 due to the symmetry of the transmittance of the Fabry-Perot cavity. The measured minimum UMSR of the comb at the HNLF input was ~65 dB at a wavelength of 1542 nm, which agreed well with the calculated value. The minimum UMSR of the comb at the HNLF output at 1542 nm was degraded to ~40 dB. The minimum UMSR at 1350 nm was also ~40 dB. It is known that the self-phase modulation induced in the HNLF reduces the UMSR [13, 14]; a degradation of 20 dB-25 dB was observed here. This is consistent with the degradation in the suppression ratio observed in the spectrum of the comb at the HNLF input and output shown in Fig. 2b. For the comb output from the cPPLN-WG, only the five signals with low UMSRs could be measured since the signal-to-noise ratios (SNRs) of the beat signals at 514 nm were low. The minimum UMSR was

~40 dB. We did not observe any significant degradation of the UMSR in the harmonic. Therefore, we believe that a UMSR level (~40 dB) similar to that in the fundamental comb is obtained at other visible wavelengths.

When comb spectra are imaged using a spectrograph, the asymmetry of the unnecessary modes on the short and long wavelength sides of the transmission modes causes shifts in the spectral positions of the optical comb modes, resulting in RV measurement errors. In particular, the asymmetry causes a significant error when the suppression ratio is low. Here, we calculated the transmitted mode spectral center-of-gravity shift [13] for a ±9th-order unnecessary-mode pair with the lowest suppression ratio (40 dB), assuming that the power difference of the mode pair was 0.5 dB (about 10 %). As a result, the estimated frequency shift was 24 kHz. This corresponds to an RV shift of 1.2 cm/s for the Doppler method in the 500 nm wavelength region, which is sufficiently small.

**Spectral stability and device durability**

First, we investigated the long-term spectral stability of the broadband 30 GHz-spacing comb in the visible range. Figure 4 shows spectra obtained at 4 h intervals over 36 h, and there was no significant change in the comb spectrum. Even after more than a year of intermittent use, there was no noticeable change in the output spectrum. In this system, much of the optical system is composed of polarization-maintaining fibers, which suppress temporal fluctuations in the spectrum due to changes in the polarization state caused by environmental changes such as variations in temperature and atmospheric pressure.

Next, we discuss the durability of the optical devices. Compared with frequency combs based on solid-state lasers such as titanium sapphire lasers, fiber-laser-based combs are robust, almost maintenance-free, and have excellent long-term functionality. In particular, Er-doped-fiber-laser-based optical combs have been widely studied [37-39]. It is also important to remember that the durability of the optical devices depends on the broadband comb generation scheme in the visible region. In this study, HNLF and the ridge-type cPPLN-WG, which are known for their high durability, are responsible for the spectral broadening and wavelength conversion of the comb, respectively. As a result, the generated visible comb power is as low as 22.4 mW for all wavelengths (360 nm-830 nm); there is a low risk of green-induced infrared absorption and other phenomena that can cause damage to nonlinear optical crystals. In fact, despite more than a year of continuous operation, the HNLF and cPPLN-WG have not needed to be replaced, and no power degradation or spectral change has been observed. We can expect the system to operate for several years without their replacement.

**Discussions**

In this section, we discuss the possibility of generating a 30 GHz-spacing comb over almost the entire visible wavelength region with minor modifications to the parameters used in the abovementioned scheme. Specifically, we assume the following changes. (1) Increase the output power of EDFA#3 to broaden the output spectrum from HNLF. (2) Extend the chirp range of the poling period of the cPPLN-WG to match the infrared spectrum of the comb output from the HNLF.

Here, we estimate how broad a spectrum can be achieved by the above improvements. Considering the available wavelength range of the third harmonics, which has the lowest power among the harmonics generated in this study (Fig. 2b), the power at the short wavelength end (453 nm) appears to be limited by the fundamental comb power (40 µW per comb mode at 1359 nm). Thus, we assumed this power per comb mode as a requirement for high harmonics generation. On the other hand, the long wavelength end (543 nm) in this work seems to be limited by the design of the cPPLN-WG, which can be improved through optimization. Figure 5 shows simulation results revealing the way in which the spectrum of the comb output from the HNLF changes when the average power of the optical pulses input to the HNLF can be increased without changing the HNLF or the spacing frequency (30 GHz) of the comb used in this study. The HNLF length was adjusted to the value where the calculated spectrum was broadest for each power. In the simulation, when the average power input to the HNLF was 3 W, the wavelength range at which the power per mode of the output comb exceeded 40 µW increased to 1279 nm-1761 nm. By designing the cPPLN-WG to satisfy the phase-matching condition in this wavelength range, the available spectral range of the second, third, and fourth harmonics were estimated to be 640 nm-881 nm, 426 nm-587 nm, and 320 nm-440 nm, respectively, which corresponds to 91 % of the frequency range in the visible wavelength region.

We realized a broadband frequency comb in the visible range based on an Er-doped fiber laser for the wavelength calibration of a high-dispersion spectrograph for astronomical observations. The mode spacing, available spectral coverage, and spectral contrast of the realized comb exceeded 30 GHz, 62 % of the visible wavelength region, and 40 dB, respectively. The results also showed excellent potential as a practical astro-comb for high-precision RV measurements, with a long-term stable spectrum, durable nonlinear optical devices, and easy frequency reproducibility using a wavelength-stabilized laser. Furthermore, simulations showed that if the comb power input into the HNLF were increased to 3 W, the spectrum of the output comb would cover 91 % of the visible wavelength region. This high-performance, easy-to-use, broadband, visible, and wide-mode-spacing comb will be a powerful tool that will encourage the widespread use of astro-combs and to take astronomical research in such fields as exoplanet exploration and the accelerated expansion of the universe to the next stage. Such combs will open the door to applications where the heterodyne-beat method is inapplicable, as well as provide existing applications with both inspiration and benefit.

## Methods

### Fully phase-stabilized Er-doped-fiber-based frequency comb

The comb source was an Er-doped-fiber-based mode-locked laser with an $f_{rep}$ of approximately 230 MHz. The output of the comb source was divided into three branches, one of which was used for visible-range comb generation as the main branch. The second branch was used to detect an $f_{CEO}$ signal with an $f$-$2f$ interferometer along with the $f_{rep}$ signal. The $f_{rep}$ and $f_{CEO}$ were phase-locked to reference frequencies based on an atomic clock by controlling the laser cavity length with an intra-cavity piezoelectric transducer and the pump power of the mode-locked laser, respectively. The third branch was used to detect the beat frequency ($f_{beat}$) between the acetylene-stabilized laser and the nearest comb mode. The configuration of the comb, except for the main branch, was essentially similar to that described in our previous work [38].

### Reference laser for locking mode-filtering cavities

In this study, we used a reference laser whose frequency matched one of the comb modes to stabilize the length of the mode-filtering cavities described below so that the cavities transmitted the comb. To obtain such a laser, we employed a 1542 nm laser frequency stabilized to an absorption line of acetylene ($^{13}C_2H_2$, $v_1 + v_3$ P(16)) [29] and feed-forward control with an in-line AOM [34]. The output of the acetylene-stabilized laser was divided into two branches, one of which was used to detect the $f_{beat}$ signal with the nearest comb mode. The output from the other branch was frequency shifted by $f_{beat}$ in the AOM so that its frequency matched the nearest comb mode. We used it as a reference laser to stabilize the optical cavity lengths so that the comb modes transmitted the cavities. The advantage of using an acetylene-stabilized laser is that the same order of the comb mode can always be used when detecting the $f_{beat}$ signal between the acetylene-stabilized laser and the comb mode. Here, we set $f_{rep}$ and $f_{CEO}$ at 230.875 909 MHz and +30 MHz, respectively. Then, from the equation $\nu(n) = n f_{rep} + f_{CEO}$, where $n$ and $\nu$ are the comb-mode order and comb-mode frequency, respectively, the frequency of the 841 879-th comb mode is approximately 194 369 609 393 kHz; a beat frequency $f_{beat}$ of approximately 40 MHz is obtained between this comb mode and the laser stabilized on the P(16) line of acetylene (194 369 569 384(5) kHz [40]).

**Mode-filtering cavities**

Three Fabry-Perot cavities were used to extract one of every 130 modes of the comb source and obtain a comb with a wide mode spacing (30 GHz) and a high UMSR. To realize this, the frequencies of the cavity transmission modes and the comb modes must be matched as precisely as possible every 30 GHz. In this study, we employed optical cavities whose FSR integer multiple matched 30 GHz. The advantage of this design is that it allows a high suppression ratio for the comb mode adjacent to the transmitted comb mode at a relatively low cavity finesse. Furthermore, by appropriately selecting the different FSRs of the three cavities, the overall UMSR could be increased. We set the FSR of each cavity at 2.00 GHz (= $130f_{rep}/15$), 1.77 GHz (= $130f_{rep}/17$), and 2.14 GHz (= $130f_{rep}/14$) in this study. In this design, 15, 17, and 14 times each FSR match the mode spacing of the transmitted comb, 30 GHz. We used the reference laser, whose frequency was matched with the nearest comb mode as described above, to stabilize the cavity FSR by controlling the cavity length so that the reference laser passed through the cavity. When stabilizing each cavity FSR, we coarsely adjusted the cavity length and selected the cavity mode in which the observed total transmitted comb power was maximized by scanning the cavity length. As a result, the FSR of each cavity was closest to the designed value (rational multiple of $f_{rep}$) and a high transmittance over a broad spectral region was obtained for the extracted comb modes. The reflectance of the cavity mirror was 97 %, which corresponded to the cavity finesse of 100. The group delay dispersion of the mirror was designed to be less than 0.15 fs$^2$ in the 1520 nm-1600 nm wavelength range, and the wavelength dependence of the FSR caused by the group delay dispersion was negligible in this range. The minimum UMSR calculated from the three cavities was 64 dB as shown by the light blue line in Fig. 4

To filter the comb mode frequencies with three optical cavities, it is necessary to obtain the desired FSRs as precisely as possible. Therefore, we have newly developed an optical cavity that does not break the beam alignment for resonance even if the cavity length is changed by a few centimeters. Two cavity mirrors are held on highly stable kinematic mounts for optical axis adjustment and installed opposite each other. One mount is directly fixed to the aluminum baseplate and the other is fixed to the baseplate via a cross-roller stage. The distance between the two mounts is basically supported by a Super Invar rod with a low thermal expansion coefficient, and a micrometer head and a piezoelectric transducer are inserted for coarse and fine tuning of the cavity length. The high linearity of the cross-roller allows the optical axis of the cavity to be maintained for resonance even when the stage is moved to change the cavity length by a few centimeters.

We employed a Pound-Drever-Hall (PDH) locking scheme [41] to stabilize the cavity FSR. To obtain the error signal for stabilization, the reference laser was phase-modulated by employing a 30 MHz sinusoidal voltage with an in-line EOM placed in front of the in-line AOM. The modulated reference laser was divided into three parts, each of which was incident in each cavity. The light reflected from the cavity was extracted by an optical circulator and incident on a photodetector, and then the cavity FSR was feedback-controlled using the error signal obtained by demodulating the detected signal.

To avoid the comb light mixing into the optics used for the PDH locking, the comb and the reference laser beams were incident in the cavity in opposite directions and with orthogonal polarizations. The comb and reference laser, respectively, were coupled to the PM fiber at the first cavity input and at the in-line AOM output through a half-wave plate, a quarter-wave plate, and a polarizer. The limited extinction ratio of a polarizing beam splitter (PBS) means that some part of the comb light amplified in EDFA#2 and EDFA#3 leaks from the PBS, and is incident on the detector used for PDH locking, thus making the PDH locking unstable. To avoid this, a narrow optical bandpass filter installed before the detector attenuated most of the comb spectrum. Furthermore, to suppress the leaked comb power relative to the reference laser power, the reference laser was amplified using EDFA#1 before the leaked comb was mixed and was attenuated to an adequate power in front of the detector.

**Chirped pulse amplification and spectral broadening**

The 30 GHz mode-spacing comb in the 1550 nm wavelength region was amplified with EDFAs and spectrally broadened with an HNLF. To obtain optical pulses with the peak power required for spectral broadening, we employed chirp-pulse amplification [35] as shown in Fig. 1a. First, the pulses were chirped with an NDF. Two EDFAs were employed and one was placed between each of the three cavities to amplify the optical power after EDFA#3 to 1.6 W. Although the pulse chirp induced by the NDF was gradually compensated for by the PM-SMF and PM-EDF used in this stage, the chirp was large until it passed through Cavity#3, and no significant nonlinear effect was observed in the amplification process. After Cavity#3, the pulse chirp was compensated to make it chirp-free using a PM-SMF with a length of about 3 m. The temporal width of the compressed pulse measured by FROG was 180 fs. From the average power of 880 mW, the pulse peak power was estimated to be 0.13 kW. The compressed pulse was incident into the HNLF with a length of 205 cm to broaden the 30 GHz-spacing comb spectrum in the near-infrared region.

**Wavelength conversion to visible region**

The comb spectrally broadened with the HNLF was incident in the ridge-type cPPLN-WG. The cPPLN-WG had a poling period that varied linearly from 12.8 μm to 19.4 μm, corresponding to the quasi-phase-matching condition of the second harmonic generation from the wavelength range 1350 nm-1600 nm to 675 nm-800 nm. The cross-section of the waveguide was a rectangle 7.8 μm high and 9.5 μm wide, and the length along the optical axis was 10 mm. Using the cascaded higher harmonic generation in the PPLN waveguide [31-33], we expected the wavelength of the comb spectrum to be converted to 675 nm-800 nm (second harmonics), 450 nm-533 nm (third harmonics), and 338 nm-400 nm (fourth harmonics).

**Spectral broadening simulation**

We simulated spectral broadening by the split-step Fourier method, which is widely used to calculate the optical pulse evolution through optical fibers [42]. The pulse waveform input into the HNLF measured by FROG was used as the initial condition for the calculation. In Fig. 2a, we set the repetition frequency and average power at 30 GHz and a measured value of 880 mW, respectively. In Fig. 5, we used average powers of 2 W and 3 W. The group velocity dispersion and nonlinear coefficient of the HNLF were −0.0022 $ps^2$/m and 20 /(W km) at a wavelength of 1550 nm. The wavelength dependence of the group velocity dispersion and nonlinear coefficient was taken into account in the calculation, and the calculated spectrum was in good agreement with the experimental result as shown in Fig. 2a. Therefore, if the average power can be increased to 2 W or 3 W without significantly changing the pulse shape, we expect a broader near-infrared comb spectrum to be obtained as shown in Fig. 5 using the same type of HNLF as discussed in the Discussion section.

**Unnecessary-mode suppression ratio measurement**

It is difficult to measure the UMSR even when using a high-resolution OSA. In fact, only the strong unnecessary modes were observed between the comb modes with a 30 GHz mode spacing in the 1550 nm region as shown in Fig. 3b. The UMSRs near the transmitted comb modes cannot be obtained correctly from this spectrum because the modes of a comb source with a 230 MHz spacing cannot be resolved with the OSA. In addition, no unnecessary modes were observed even when Fig. 3d was presented in a logarithmic scale (not shown here) because of the lower frequency resolution of the OSA in the 800 nm region.

In this study, we observed heterodyne beats for UMSR measurement. We first measured the SNR of the beat between a CW laser and the comb transmitted mode as the reference SNR and then measured the SNR of the beats between the CW laser and unnecessary comb modes. The ratio between them is the UMSR. If the CW laser power is sufficiently larger than the total comb power including the mode power $P$ and the noise equivalent power of the photodetector, the SNR of the heterodyne beat is expressed as $P/(2h\nu\Delta f)$ and does not depend on the CW laser power, where $h$ is the Planck constant, $\nu$ is the optical frequency, and $\Delta f$ is the resolution bandwidth. Therefore, even if the CW laser power changes when the beat signal with each comb mode is measured by changing the CW laser frequency, it has little effect on the UMSR measurement results. Using this approach, we measured the UMSRs of the comb at the HNLF input, at the HNLF output, and at the cPPLN-WG output. For the SNR measurement at the HNLF input and output, we offset-locked a 1542 nm CW laser to each mode of the 230 MHz-spacing comb source and measured the SNRs of the beat signals between the CW laser and the transmitted and suppressed modes of the 30 GHz-spacing comb at 1542 nm. For the evaluation at the cPPLN-WG output, the 1542 nm CW laser, which was offset-locked to the comb source before passing through the cavities, was wavelength-converted to 514 nm by third harmonic generation with a dual-pitch PPLN waveguide [43], and beat signals with the 30 GHz-spacing comb at a wavelength of 514 nm were observed. We measured the SNRs of the beat signals by offset-locking the CW laser to the transmitted and suppressed modes of the comb source as in the evaluation at 1542 nm and determined the UMSRs by calculating their ratio. The offset locking was performed with a control bandwidth of several hundred kHz so that the linewidth of the coherent peak of the in-loop beat spectrum was sufficiently narrow. This enabled us to observe the beat signal at a low resolution bandwidth with a high SNR, and thus made it possible to measure a high UMSR. On the other hand, for a beat signal with a high SNR, it was difficult to distinguish the noise floor from the sideband components around the beat frequency, which was an uncertainty factor in the SNR measurement and thus also in the UMSR measurement. In addition, we considered the frequency response and saturation characteristics of the photodetector, the power fluctuations of the comb, and the measurement uncertainty of the RF spectrum analyzer as uncertainty factors, and we concluded that the measurement uncertainty of the UMSR was several dB.

## Acknowledgments

We would like to thank Dr. Nishida of NTT Electronics for assisting with the design of the cPPLN waveguide used in this study and Prof. Hong and Dr. Ikeda of Yokohama National University for aiding the evaluation of the cPPLN waveguide. We are also grateful to Dr. Izumiura and Dr. Kambe of the National Astronomical Observatory of Japan for helpful discussions as regards the required specifications for the comb used for astronomical radial velocity measurement. This work is supported by JST ERATO Grant Number JPMJER1304 and JSPS KAKENHI Grant Number 15K21733.

## Author contributions

All authors designed the system, discussed the results, and wrote the manuscript; K.N. was responsible for most of the experimental setup and measurements with assistance from K.K., S.O. and H. I; K.K. was responsible for simulating the spectral broadening and constructed part of the experimental setup; S.O. was responsible for the electronics system; H.I. led the project.

## Competing interests

The authors declare no competing interests.

# Figures

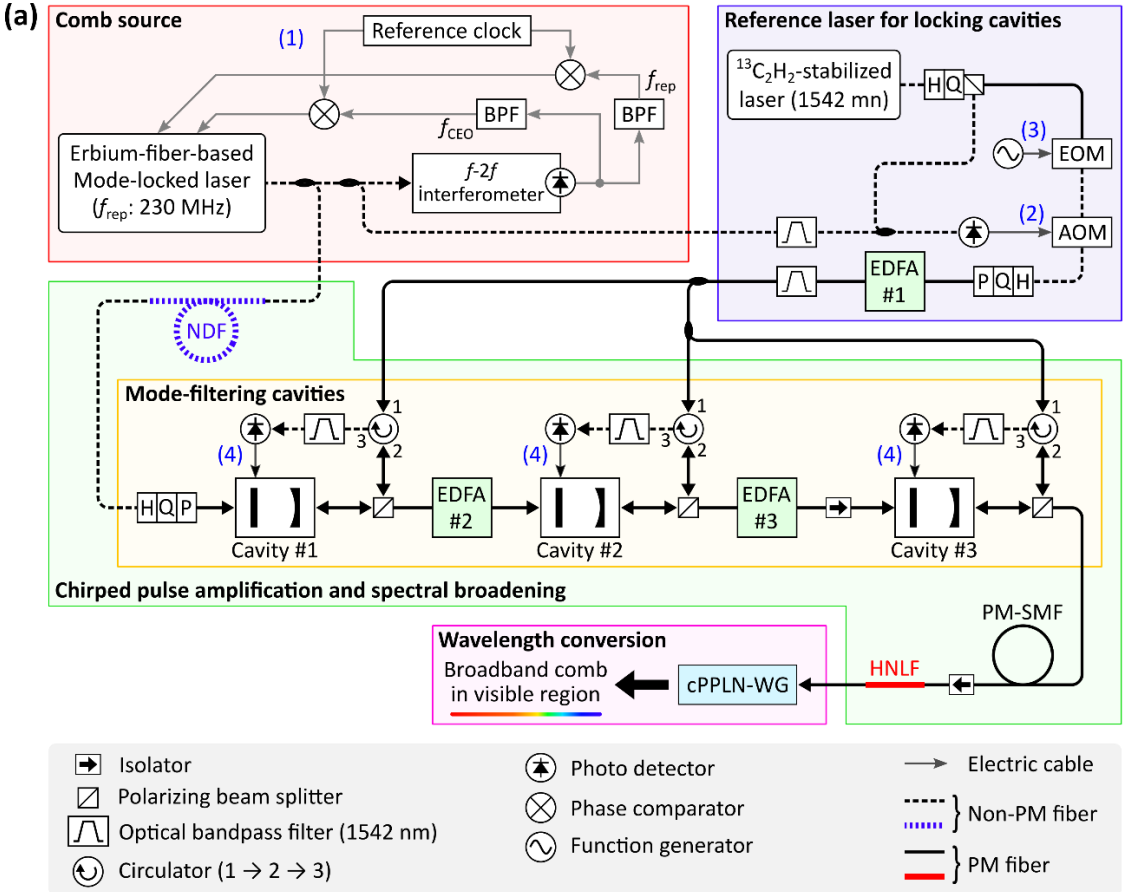

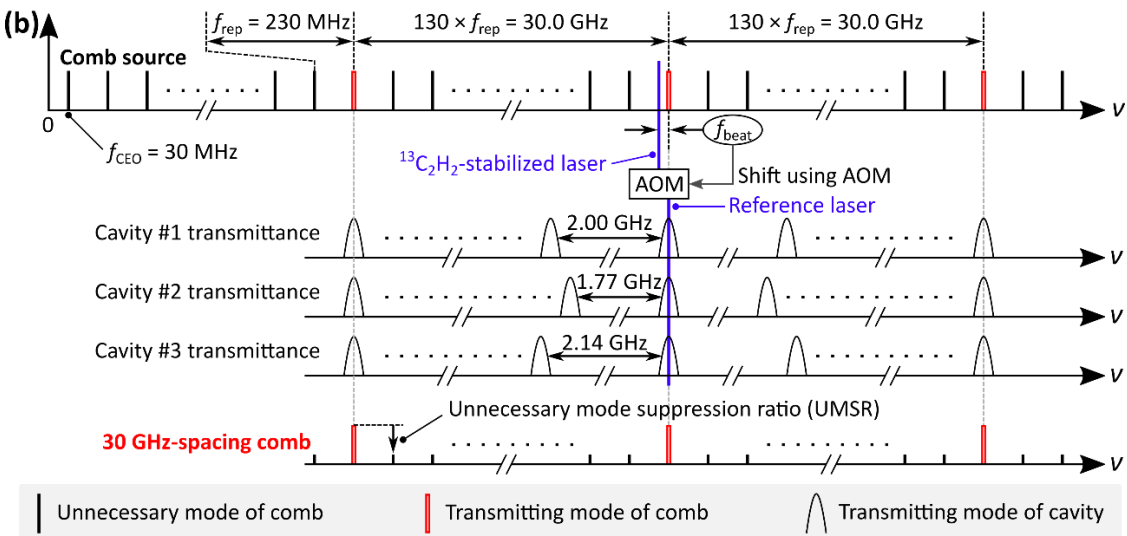

**Figure 1 | Experimental schematic of an Er-fiber-based visible-broad frequency comb with a 30 GHz mode-spacing. a,** Schematic of the comb system. (1) $f_{rep}$ and $f_{CEO}$ are stabilized to reference frequencies by controlling an intracavity piezoelectric-transducer and the pump power for the mode-locked laser, respectively. (2) The beat note between the acetylene-stabilized laser and the comb is fed forward to an acousto-optic modulator (AOM) to match the frequency-shifted acetylene-stabilized laser to the comb-mode frequency. We call this frequency-shifted light a reference laser. (3) The acetylene-stabilized laser is phase-modulated with an electro-optic modulator (EOM). (4) The lengths of three mode-filtering cavities are stabilized to the frequency-shifted acetylene-stabilized laser with a Pound-Drever-Hall scheme. BPF, electrical bandpass filter, EDFA, Er-doped-fiber amplifier, H, half-wave plate, Q, quarter-wave plate, P, polarizer, PM, polarization-maintaining, NDF, normal-dispersion fiber, SMF, single-mode fiber, HNLF, highly non-linear fiber, cPPLN-WG, chirped-periodically-poled lithium niobite waveguide. **b,** Frequency relation between the comb modes, the transmission modes of the mode-filtering cavities, and the reference laser in the optical frequency domain.

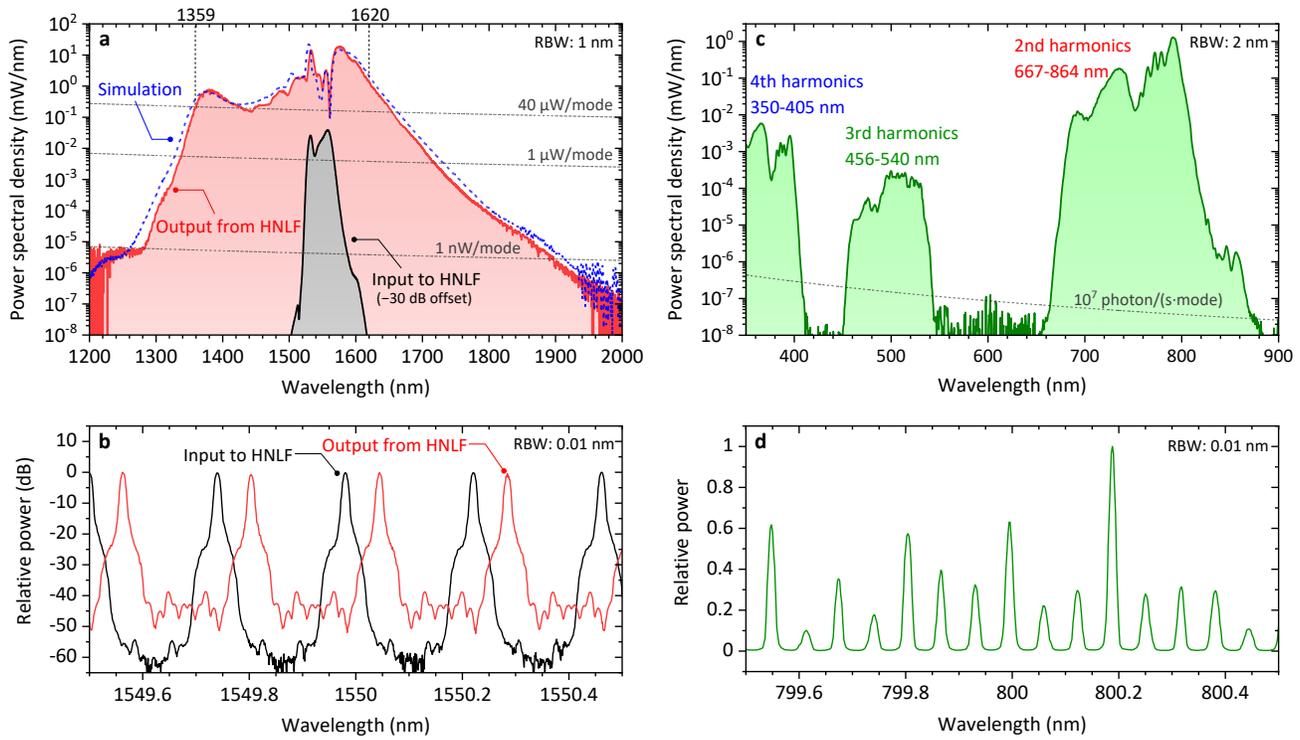

**Figure 2 | Comb spectrum. a,** Comb spectrum in the near-infrared region. The solid black line shows the spectrum of the comb at the HNLF input, shifted by −30 dB. The solid red and dashed blue lines, respectively, show the measured and simulated spectra of the comb at the HNLF output. **b,** Comb-resolved spectrum at the input (black line) and output (red line) of the HNLF. The difference between the center wavelengths of each mode of the two spectra is due to drift in the instrument (OSA). **c,** Spectrum of the comb at the cPPLN-WG output in the visible region. The dotted black line corresponds to $10^7$ photons per comb mode, which is a sufficient number of photons to be used as the wavelength reference by the high-dispersion spectrograph. **d,** Comb-resolved spectrum at the output from a cPPLN-WG.

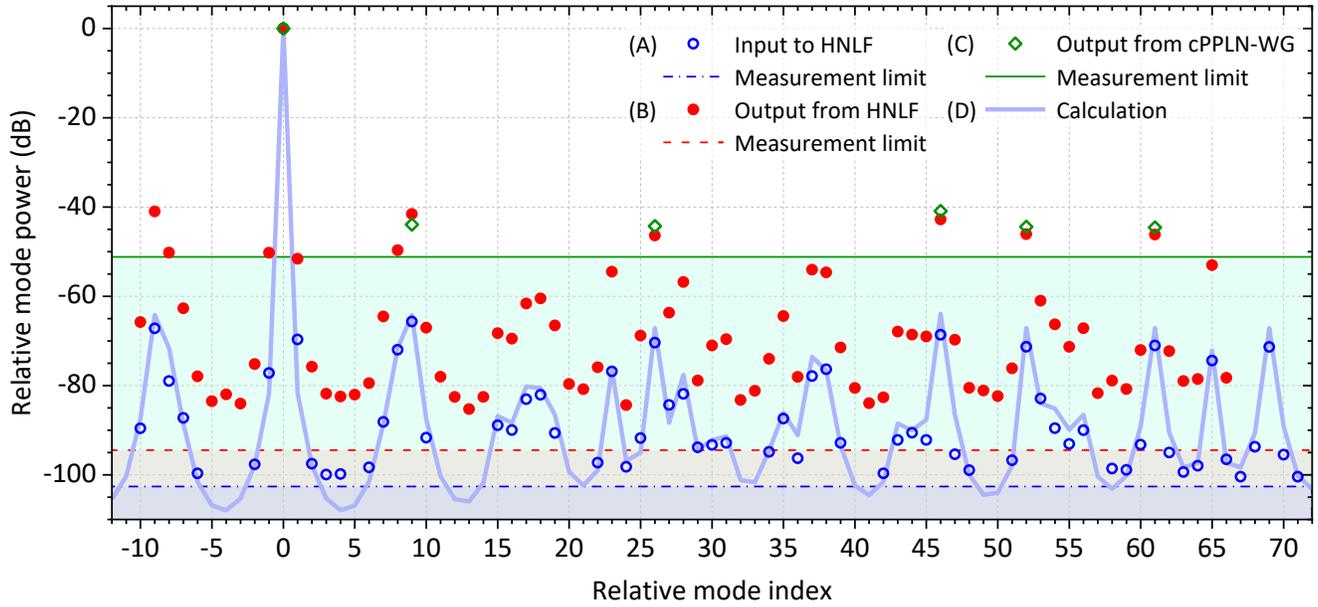

**Figure 3 | Unnecessary-mode suppression ratio of infrared and visible combs.** Unnecessary-mode suppression ratio (UMSR) measured from the SNR of the beat signals between a probe laser and the modes of (A) comb input to the HNLF (1542 nm, open blue circles), (B) comb output from the HNLF (1542 nm, filled red circles), and (C) comb output from the cPPLN-WG (514 nm, green diamonds). A continuous-wave laser (1542 nm or 514 nm) was used as the probe by sweeping a range of ~20 GHz. The blue chain, dashed red, and solid green lines show the measurement limits of the UMSRs for (A), (B), and (C), respectively, which are limited by the SNR of each beat signal. (D) The light blue line shows the transmittances of comb modes calculated from the designed finesses of the three optical cavities and the ratios of the FSRs of the cavities to the comb's $f_{rep}$.

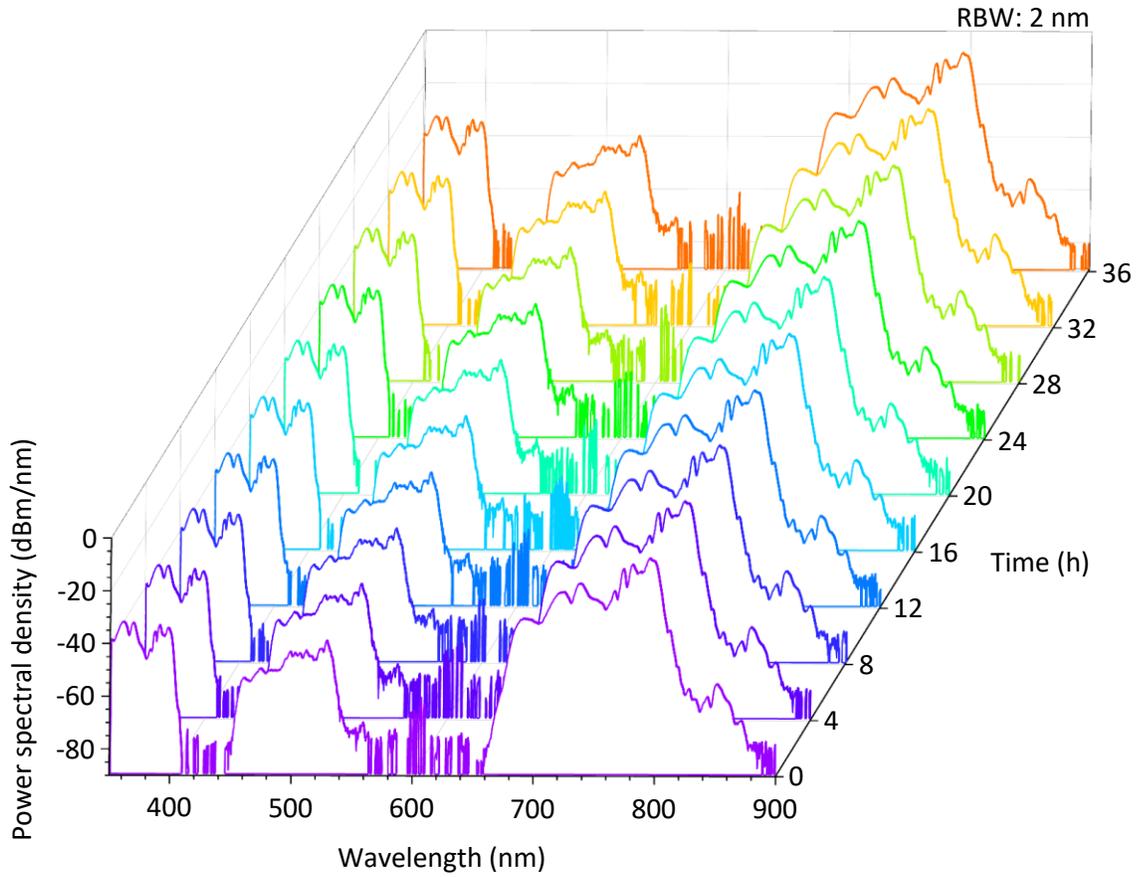

**Figure 4 | Long-term stability of comb spectrum.** Spectrum of a broadband comb in the visible range with a mode-spacing of 30 GHz recorded every 4 hours for 36 hours. The spectral shape is almost unchanged.

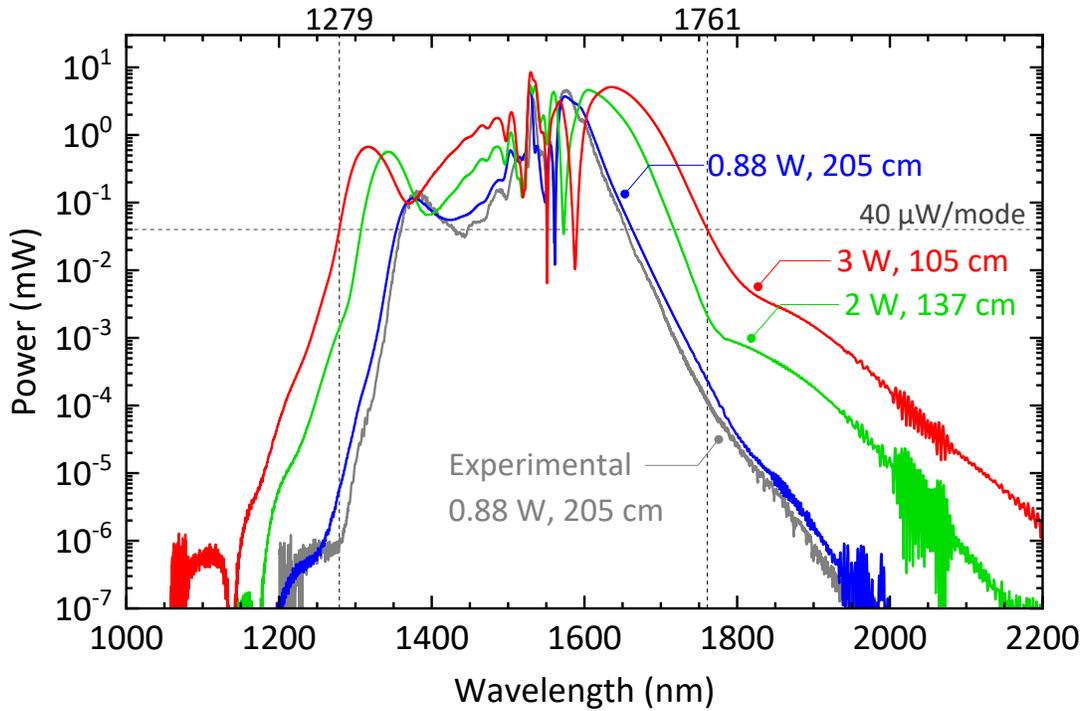

**Figure 5 | Measured and simulated spectra of comb output from HNLF.** The current spectrum of the broadband comb in the near-infrared range and the simulated spectra that would be obtained were the output power from the final-stage EDFA to be enhanced. The HNLF length was optimized for each condition.